Finding New Superconductors: The Spin-Fluctuation Gateway to High $T_c$ and Possible Room Temperature Superconductivity

David Pines, Physics Department, UC Davis, ICAM, UIUC, and the Santa Fe Institute

Abstract

We propose an experiment-based strategy for finding new high transition temperature superconductors that is based on the well-established spin fluctuation magnetic gateway to superconductivity in which the attractive quasiparticle interaction needed for superconductivity comes from their coupling to dynamical spin fluctuations originating in the proximity of the material to an antiferromagnetic state. We show how lessons learned by combining the results of almost three decades of intensive experimental and theoretical study of the cuprates with those found in the decade-long study of a strikingly similar family of unconventional heavy electron superconductors, the 115 materials, can prove helpful in carrying out that search. We conclude that since $T_c$ in these materials scales approximately with the strength of the interaction, J, between the nearest neighbor local moments in their parent antiferromagnetic state, there may not be a magnetic ceiling that would prevent one from discovering a room temperature superconductor.

Keywords

Unconventional superconductors
Heavy electron superconductors
Cuprate superconductors
Emergent superconductivity

Introduction

Finding new classes of materials in which superconductivity emerges at temperatures comparable to, or greater than, the highest transition temperature achieved in the cuprate superconductors is one of the major scientific challenges of our new century. It is rendered difficult, in part, by the fact that superconductivity is a poster child for the kind of emergent behavior in quantum matter that is more easily discovered in the laboratory rather than using theory as a guide. In this paper we ask to what extent the lessons learned from the intensive experimental and theoretical study of the cuprates for almost three decades and the decade-long study of a strikingly similar family of unconventional heavy electron superconductors, the 115 materials, can prove helpful in carrying out that search.

We begin by describing our proposed search strategy. It is based on the by now well-established spin fluctuation gateway to superconductivity in which the attractive quasiparticle interaction needed for superconductivity comes from their magnetic coupling to dynamical spin fluctuations originating in the proximity of the material to an antiferromagnetic state [1][2]. We next present the arguments for pursuing this strategy that emerge when one combines experiment, phenomenology, and model calculations for the cuprate and 115 superconductors. We conclude with a discussion of the possibility of finding a room temperature superconductor.

A Suggested Strategy

Our proposed experiment-based strategy for discovering new families of high $T_c$ superconductors can be simply stated:

* search for new classes of antiferromagnetic materials whose properties can be changed by doping and/or pressure

*follow their dynamic magnetic properties closely and search for signs of nearly two-dimensional magnetic and quasiparticle behavior

*search for materials containing local moments whose nearest neighbor interaction J, is at least comparable to, but preferably considerably larger, than that found in the cuprates

*Because chemists are good at making stuff, if you are not a chemist, find a chemist as a collaborator

The first two "rules" are based on what experiment, phenomenology, and model calculations have taught us about the cuprates. The third emerges when one compares the results of NMR experiments on the 115 family of heavy electrons, whose properties are remarkably similar to the cuprates, with those on the cuprates. The fourth is proposed, in part, to encourage more chemists to join physicists in the search for new superconductors.

Lessons Learned from the Magnetic Behavior of the Cuprates

An overview of the remarkable properties of cuprate superconductors such as $YBa_2Cu_3O_{6+x}$ [the 123 materials] and

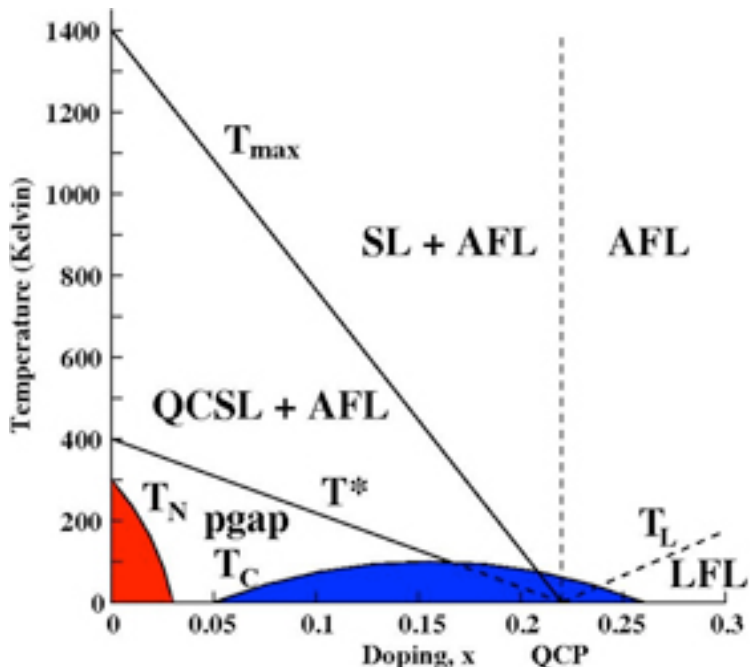

Figure 1: A candidate phase diagram for the cuprate superconductors based, in large part, on magnetic measurements of normal state behavior. It depicts the changes in their emergent behavior and ordering temperatures as a function of the concentration of holes in the CuO planes. The red and blue shaded regions are those in which one finds, respectively, local moment antiferromagnetic order, and quasiparticle superconducting d-wave order. $T_{max}$ is the temperature at which the temperature-dependent uniform static spin susceptibility reaches its maximum value, $T^*$ is the temperature at which the pseudogap phase emerges, $T_N$ is the Neel transition temperature for local moment AF ordering, $T_C$ is the quasiparticle superconducting transition temperature, $T_L$ is the temperature below which the quasiparticle Fermi liquid (FL) displays Landau FL behavior (which will appear in the superconducting region if magnetic fields strong

enough to destroy superconductivity are applied) while above $T_L$ (and T*) one finds the anomalous Fermi liquid (AFL) transport behavior expected for quasiparticles being scattered against quantum critical fluctuations emanating from the upper quantum critical point, shown here at a doping level ~0.21, although its precise location may be at a somewhat lower doping level. To the left of this QCP, local moments and quasiparticles co-exist over the entire region of doping (x < 0.21), while to its right, no local moments are present below a temperature that has not yet been determined. Above $T_{max}$ the local moments form a spin liquid that exhibits the mean field scaling behavior expected for a two-dimensional Heisenberg antiferromagnet at high temperatures, while below it they display the behavior expected for a two-dimensional Heisenberg antiferromagnet in the quantum critical regime, hence the notation, quantum critical spin liquid (QCSL). Not shown on the figure is the low doping quantum critical point. From D.Pines, Unit 8 in "Physics for the Twenty-first Century", http://physics.digitalgizmo.com/courses/physics/unit/text.html?unit=8&secNum=0

$La_{2-p} Sr_p CuO_4$ [the 214 materials] is given in the caption to the phase diagram, Figure 1, that shows the consequences of adding holes in their Cu-O planes through chemical substitution. At zero hole doping the planar Cu quasiparticles are localized by their strong Coulomb repulsion and order antiferromagnetically. As holes are added, the low temperature long range AF order disappears and an unconventional superconducting state, in which the planar quasiparticles pair in a d-wave state, emerges.

Before the 1994 direct determination of d-wave pairing in the cuprates in phase sensitive tunneling experiments, there were a number of reasons to believe that the magnetic glue responsible for their superconductivity might be the dynamical spin fluctuations associated with their proximity to an antiferromagnetic state. Thus shortly after the discovery of the cuprate superconductors, some of us who had been thinking about spin-fluctuation induced superconductivity in heavy electron materials put forth the idea that a similar mechanism might be responsible for high $T_c$ in the cuprates [3]. When the proposal was discussed further at the 1989 Los Alamos Symposium, it aroused a good deal of controversy [4], although measurements of the Knight shift in the superconducting state of optimally doped YBCO reported there showed a temperature dependence consistent with d-wave pairing, while the quite different spin-lattice relaxation rates measured for planar $^{63}$Cu and $^{17}$O nuclei provided an essential clue to the dynamical spin fluctuation spectrum, since this difference had been explained using a two-dimensional spin fluctuation spectrum that was peaked at the commensurate wavevector, Q= ($\pi,\pi$), and displayed quantum critical behavior [5]

$$\chi(q,\omega) = \chi_Q /[1 + (Q-q)^2\xi^2] - i\omega/\omega_{sf}] \qquad [1]$$

where $\chi_Q = \alpha\xi^2$ is the commensurate static spin susceptibility, and $\xi$ is the antiferromagnetic correlation length.

Since the coupling of quasiparticles to dynamical spin fluctuations leads to an effective quasiparticle interaction that is proportional to $\chi(q,\omega)$, and the latter can become quite large as $\xi$ increases with decreasing temperature, such spin fluctuations represented an attractive candidate for the physical origin of superconductivity in the cuprates, so the idea deserved further investigation despite the fact that earlier calculations of their effectiveness in bringing about superconductivity using a single band Hubbard model had not led to promising results

Detailed independent calculations of the effectiveness of this candidate magnetic glue in bringing about high temperature superconductivity were then carried out by research groups in Tokyo [6] and Urbana [7], who found in both weak and strong coupling calculations that the pairing state had to be d-wave and that $T_c$ could be ~100K for a reasonable choice of parameters. The Tokyo group [6] used a quasiparticle interaction taken from Moriya's self-consistent renormalization [SCR] theory of quantum critical fluctuations. The Urbana group [7] used novel computational techniques developed by Monthoux to calculate the full consequences of an effective quasiparticle interaction that was proportional to the dynamic magnetic susceptibility, Eq.1, with parameters chosen to fit the NMR experiments. The Urbana group found that taking into account explicitly the wavevector and frequency dependence of this interaction led to superconducting transition temperatures that were almost an order of magnitude larger than those obtained by averaging over its consequences, and that for quasiparticles with a Fermi surface similar to that proposed for the cuprate superconductors [and in agreement with the Tokyo group] d-wave pairing was an inevitable consequence of this spin-fluctuation induced interaction.

The 1994 experimental confirmation [8] of the cuprate pairing state predicted by the Tokyo and Urbana groups obviously strengthened the case for its spin-fluctuation origin, but it did not bring an immediate "occam's razor" moment in which most members of the condensed matter community accepted their proposed spin-fluctuation gateway to high temperature superconductivity. There were a number of reasons for this. First, "complexity"-- their work did not explain every aspect of the phase diagram shown in Fig.1, including the role played by the pseudogap behavior seen in the underdoped YBCO materials and and the closely related issue of the change of $T_c$ with doping, so other approaches might be needed to attack the full phase diagram. Second, "not invented here"-- a reluctance by theorists who had developed other approaches to accept the possibility that the origin of high $T_c$ in the cuprates had been resolved using an approach that was different from their own. Third, "funding fear"--a reluctance by some experimentalists to agree that a key issue about the cuprates had been resolved because they thought that if they did so they might lose their funding for continued work on these fascinating materials

Some twenty years later, these concerns have not completely abated. It therefore seems worthwhile to address the first group of concerns in the context of our current experiment-based phenomenological understanding of the interplay between quasiparticle localization, itinerancy, spin fluctuations, and unconventional superconductivity that produces the phase diagram shown in Figure 1. Doing so is useful as well because the physical picture that has emerged is at variance with the conventional "wisdom" that has informed so many toy models, while reminding the reader that a full microscopic description of that interplay Is yet to be developed.

For our purposes--proposing a strategy to find new superconductors--that phenomenological understanding of the phase diagram will suffice. It is based on spin forensics-- using the scaling behavior of the temperature-dependent uniform magnetic susceptibility [shown in Fig. 2], whose origin was identified as that of a 2-d Heisenberg antiferromagnet, in work by Johnston [9] that was subsequently extended by Nakano et al [10], to establish the co-existence of two fluids-- a two-dimensional local moment Heisenberg spin liquid and a quasiparticle fermion liquid-- throughout a doping region that includes the optimal doping at which $T_c$ is maximum. This technique made it possible for Barzykin and the writer [11], hereafter BP, to obtain the dynamical scaling results shown in Fig.3, and to explain the measured static spin susceptibility of the cuprates in the normal state with a simple two-fluid expression:

$$\chi(p) = f(p)\chi_{FL} + [1-f(p)]\chi_{SL}(p,T) \qquad [2]$$

where p is the hole doping level, and f(p) is a doping-dependent order parameter describing the relative strength of the itinerant quasiparticles that emerge through d-p hybridization as the material is doped. As BP note, in contrast to the two-fluid model for heavy electron materials that inspired their approach (and is discussed in detail below) the order parameter describing the itinerant quasiparticle emergence is temperature independent, because the hybridization occurs at temperatures [~eV] large compared to those in which we are interested.

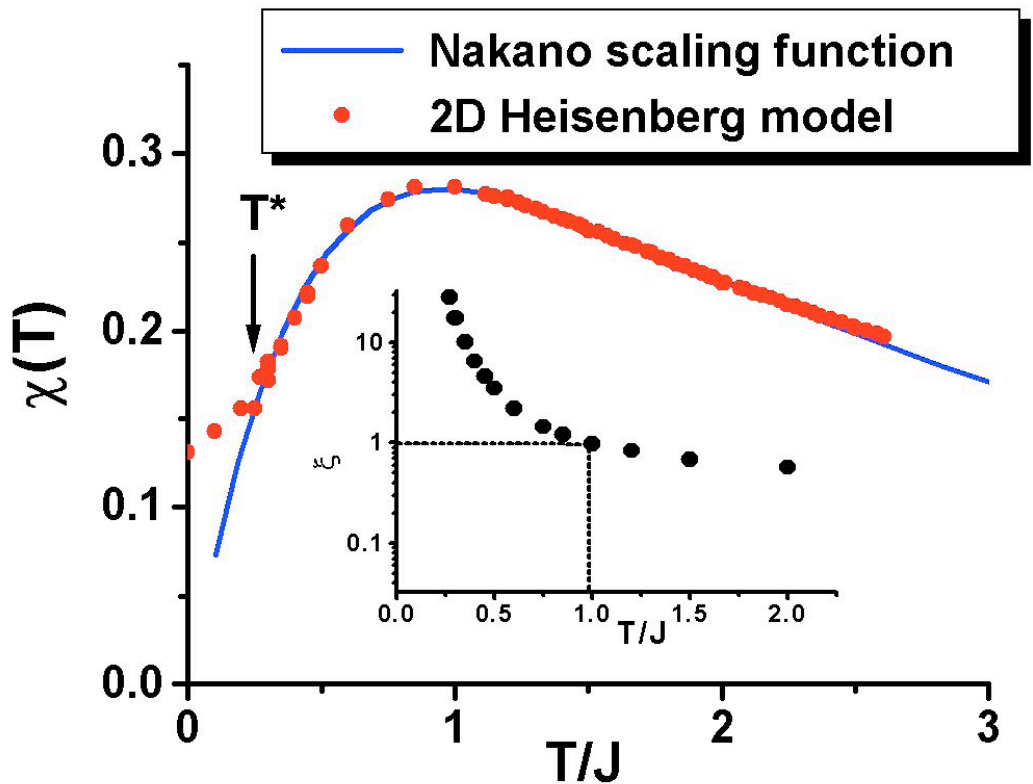

Figure 2. Comparison of the Nakano scaling function [10] for the bulk spin susceptibility data in underdoped metallic La$_{2-x}$Sr$_x$CuO$_4$ to the Heisenberg model numerical calculations of Makivic and Ding [12]. The maximum for spin susceptibility is reached at $T_m \sim 0.93J$. Deviations from the Heisenberg model results are observed for $T < T^* \sim J/3$. The inset shows the numerical results for the correlation length [11] that demonstrate that $\xi \sim 1$ at temperature $T^* \sim T_m$. Figure from BP [11].

Experiment shows that the static quasiparticle spin susceptibility $\chi_{FL}$ is temperature independent, as might be expected for a Landau Fermi liquid or, rather more likely, a two-dimensional electron glass, while, as Johnston and Nakano et al emphasized, the temperature and doping dependent $\chi_{SL}(p,T)$ is that calculated for a 2d Heisenberg spin liquid [12], with a doping-dependent effective interaction between nearest neighbors given by

$$J_{eff}[p] = [1-f(p)] \, J[0]. \quad [3]$$

The observed maximum in $\chi_{SL}(p,T)$ occurs at $T_m[p] = 0.93 J_{eff}(p)$.

BP established the doping dependence of the order parameter, f, from measurements of the doping dependence of $T_m(p)$; they find for $La_{2-p}Sr_pCuO_4$

$$[1-f(p)] \sim [1- 4.76p], \quad [4a]$$

while for $YBa_2Cu_3O_{6+x}$ the doping level is measured by x and one has the approximate result,

$$[1-f(x)] \sim [1- 0.8x]. \quad [4b]$$

As may be seen in Fig. 2, a comparison of numerical calculations of Heisenberg behavior [12] with the Johnson/Nakano fits to $\chi_{SL}$ makes it possible to identify T* as the emergence of pseudogap behavior in the Heisenberg spin liquid; importantly, because the spin liquid behavior below T* also scales with J, the physical origin of the pseudogap must be the nearest neighbor local moment interaction. This conclusion is consistent with the BP finding that $T^*[p] \sim J_{eff}[p]/3$ for the 123 and 214 materials.

Moreover, the BP analysis provides magnetic evidence for the existence of two quantum critical points [QCP] in the cuprates. As may be seen in Fig. 3, the first QCP is identified from the doping dependence of the off-set, A[x] in the $Cu^{63}$ spin-lattice relaxation rate measurements of $T_1$, for which $T_1T \sim [A[x] + BT] \sim [T_0(x) + T]$; it occurs at low dopings and marks the onset of long-range antiferromagnetic order in the spin liquid. The strongly momentum-dependent QC spin fluctuations to which it gives rise are described by Eq [1] and are responsible for the spin-lattice relaxation rate of $^{63}Cu$ nuclei at doping levels >0.05 for the 214 materials, > 0.33 for the 123 materials.

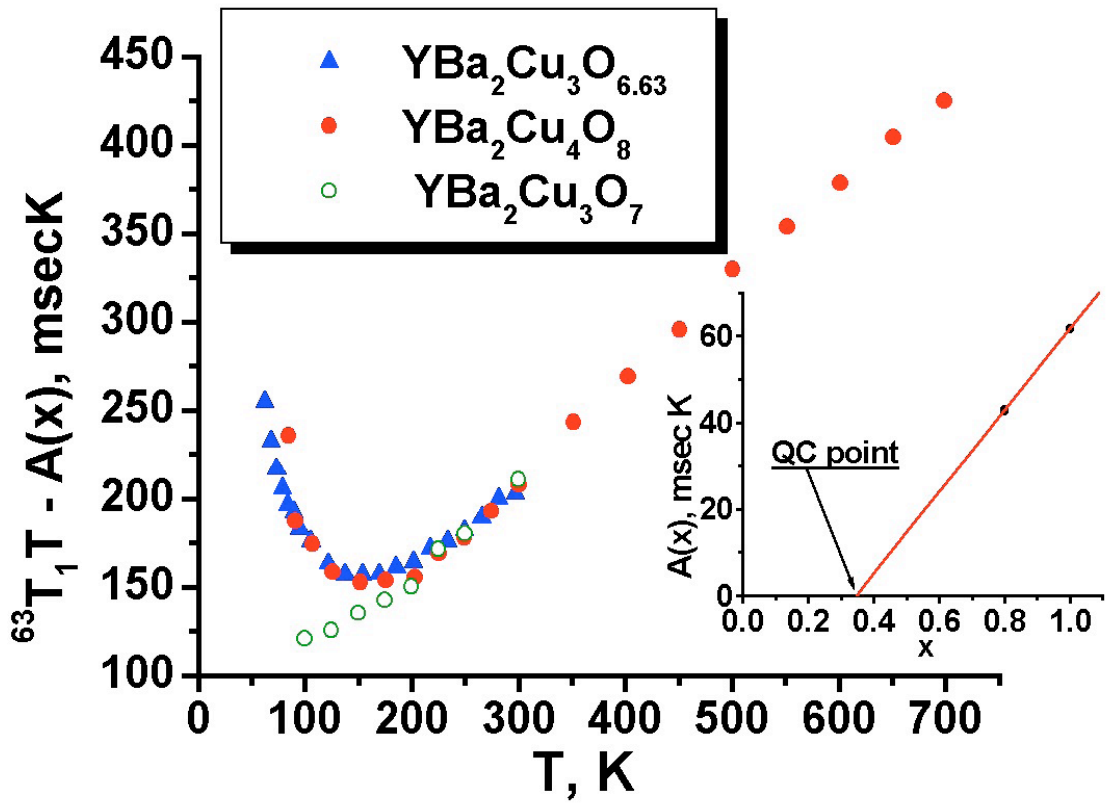

Figure 3. Scaling for the NMR relaxation rate $T_1$ of the spin liquid in $YBa_2Cu_3O_{6+x}$. Note the departures from linear in T behavior that signal the emergence of pseudogap behavior at T* [~120K for $YBa_2Cu_3O_7$ and significantly higher for the other materials]. The offset $A(x)$ shown in the inset depends linearly on x, and points to a QCP in the spin liquid at x~0.33. Figure reproduced from BP [11]

The second QCP marks the emergence of quasiparticle localization and spin liquid behavior at T=0 as one decreases doping from the overdoped side of the phase diagram. For the 214 materials, the BP analysis predicts that this second QCP will be located near p~0.21; for the 123 materials, it is expected to be located near x=1.25. It is reasonable to argue that it is the competition of quasiparticle localization with their superconductivity that prevents the cuprates from achieving much higher superconducting transition temperatures, since, as one starts from the highly over-doped side, and decreases doping, $T_C$ would appear poised to continue to increase with reduced doping; localization, however, intervenes to bring about a maximum and subsequent reduction in $T_C$. Viewed from an underdoped perspective, the growth of $T_C$ with increased doping, identified by Uemura et al [13] as being proportional to the superfluid density, is a direct reflection of the incomplete localization measured by f, and scales with f until one reaches near-optimum doping levels.

It is tempting to speculate that it is the QC fluctuations associated with this second QCP that give rise to the marginal Fermi liquid [MFL] behavior seen in resistivity and other measurements, since one would expect their fluctuation spectrum to be spread uniformly in momentum space as has been proposed for the MFL [14]. If this speculation is correct, then although these fluctuations give rise to the MFL quasiparticle behavior, they are not the magnetic glue responsible for their superconductivity, since a d-wave pairing state induced by spin fluctuations requires a fluctuation spectrum that is peaked at or near the commensurate wavevector.

Subsequent NMR experiments confirmed the BP two fluid model and established the changes in the behavior of the two fluids in the superconducting state. In three seminal papers, Haase, Slichter and their collaborators find that both $\chi_{FL}$ and $\chi_{SL}$ are present in the Knight shift of the following cuprates: 214 at 0.15 doping [15]; $YBa_2Cu_4O_8$ under pressure [16]; and under-doped and optimally doped $HgBa_2CuO_{4+\delta}$ [17]. Their measurements confirm that $\chi_{FL}$ is temperature independent above $T_C$, while below $T_C$ it reflects the expected d-wave superconductivity. This behavior is in sharp contrast to the temperature-dependent $\chi_{SL}$ that does not change character when the material becomes superconducting.

Since superconductivity arises out of the quasiparticle component (whose temperature-independent spin susceptibility, given the inhomogenous behavior of the underdoped cuprates is likely that of a 2-d electronic glass, dubbed electronic mayonnaise by Schmalian and Wolynes (18). the Haase/Slichter results provide strong experimental justification for extending to the underdoped region the approach developed by the Tokyo and Urbana groups for overdoped and optimally doped cuprates. Hopefully what will emerge is a qualitative, if not quantitative, fit to the measured $T_c$ dome.

A further challenging fundamental problem is that of understanding at a microscopic level the emergence of "Mottness" and pseudogap behavior as one starts from the overdoped side and reduces the doping; here, quite new theoretical approaches might be required [19].

Some revealing toy model calculations

Some fifteen years ago, Monthoux and Lonzarich began a series of simple toy model calculations that enabled them to explore the effectiveness of the magnetic mechanism and compare it with that of charge-density wave induced superconductivity. Because their results can further inform the search for new families of superconductors, their principal results are summarized here:

*magnetically-mediated superconductivity is much more effective in quasi two dimensions than in three dimensions because in two dimensions one has much better impedance matching between the nodes in the superconducting d-wave gap and regions of strong repulsion [20].

* an explicit study of the cross-over from cubic behavior to a tetragonal lattice explains why $T_c$ for the parent of the 115 materiais, the cubic lattice material, $CeIn_3$, is much smaller than that observed in the more anisotropic 115 materials [21].

*a detailed study of the difference between the possible p-wave superconductivity at low temperatures that could occur near ferromagnetism and the d-wave superconductivity that occurs near antiferromagnetism led to the first discovery of superconductivity in an almost ferromagnetic material, $UGe_2$ [22].

*band structure can play a significant role in determining both $T_c$ and the nature of the pairing state and it is the differences in band structure that explain the different pairing states and transition temperatures for the cuprates and the ruthenates [23].

* when an approach similar to that employed for spin fluctuation-induced superconductivity is used to study the nature of density-fluctuation-induced superconductivity in materials on the border of a

density instability, one finds that density-fluctuation-induced superconductivity is also more robust in quasi- two dimensions than three, and in anisotropic tetragonal lattices than in cubic, thereby providing another guide to finding new superconducting families [24].

The interested reader is referred to their papers for further details.

The 115 materials

A natural question to ask at this stage is whether the cuprates have already provided a useful guide for experimentalists in their search for other families of unconventional superconductors. Work over the past decade has shown the answer to this question is "Yes" in the case of the 115 [$CeMIn_5$ and $PuMGa_5$ where M=Co, Ir, or Rh] members of the family of heavy electron superconductors, whose crystal structure is shown in Fig.4. As we now show, the 115 heavy electron materials turn out to be remarkably similar to the cuprates. Not only do the guidelines developed from our examination of the cuprates prove applicable to these materials, but experimental results on the 115 materials enable us to expand these.

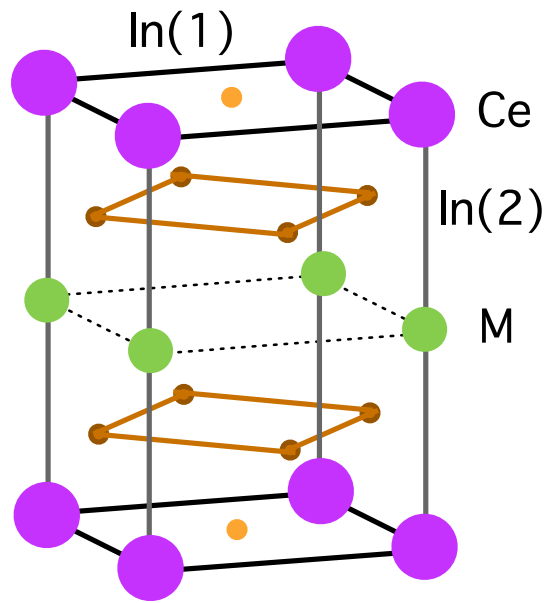

Fig, 4. Crystal structure of 115 heavy electron materials.

Quite generally, heavy electrons are found in materials that at room temperature contain a lattice of local f-electron moments coupled magnetically to a conduction electron background. As the temperature is lowered the local moments lose their substantial entropy by hybridizing with the conduction electrons, rendering some of the latter sufficiently heavy that their average mass can sometimes exceed several hundred bare electron masses. The hybridization process is gradual, beginning at what is often called the coherence temperature, T*, and is often incomplete when superconductivity emerges at low temperatures. Thus the heavy electrons become superconducting in the presence of a lattice of incompletely hybridized local moments, a two fluid situation [25] quite like that found in the cuprates.

For most, and perhaps all, heavy electron materials, the hybridization process is not via the single ion local moment screening known as the Kondo effect; the latter is characterized by the single ion Kondo temperature, $T_K \sim [\exp -1/J^*\rho]/\rho$, where $J^*$ is the strength of its Kondo coupling to the background conduction electrons whose density of states is $\rho$. Instead, the hybridization is collective and the energy scale of its onset, the coherence temperature, T*, is quite large compared to $T_K$ [26].

Importantly, the emergent heavy electron liquid is a new state of matter, called a Kondo liquid; it displays universal behavior that is not that of a Landau Fermi liquid, but instead scales with T* over a wide range of temperatures between T* and its eventual ordering temperature [27]. As befits collective hybridization, T* is determined by the nearest neighbor local moment interaction induced by their Kondo coupling to the conduction electrons; thus

$$T^* = cJ^{*2} \rho >> T_K$$

where c is a constant determined by the details of the hybridization and the conduction electron Fermi surface, [28].

A two fluid model description of the magnetic behavior of the emergent heavy electron liquid and the hybridized lattice of local moments was developed to explain the emergence of heavy electron behavior in the 115 materials [25] several years prior to the BP description of the cuprates. In it, the uniform magnetic susceptibility, $\chi$, takes a form similar to Eq. 2,

$$\chi(P,T) = f[T,P]\chi_{KL} + [1-f\{P,T\}]\chi_{SL} \qquad [5]$$

where P denotes pressure, and $\chi_{KL}$ and $\chi_{SL}$ are the susceptibilities of the Kondo liquid and the hybridized f-electron local moment spin liquid respectively. The pressure and temperature dependent Kondo liquid order parameter, f[P,T], describes the emergence of the Kondo liquid as a result of a collective hybridization of the local moments with the background conduction electron; it was subsequently [27] found to be

$$f[P,T] = f_0[P][1-T/T^*]^{3/2} \qquad [6]$$

This two fluid description provides the basis for a new phase diagram for heavy electron materials that is shown in Fig.5; in this replacement for the familiar, but no longer valid, Doniach diagram. the hybridization effectiveness parameter, $f_0[P]$, may be seen to place significant constraints on the nature of the emergent ordered states at low temperatures [28].

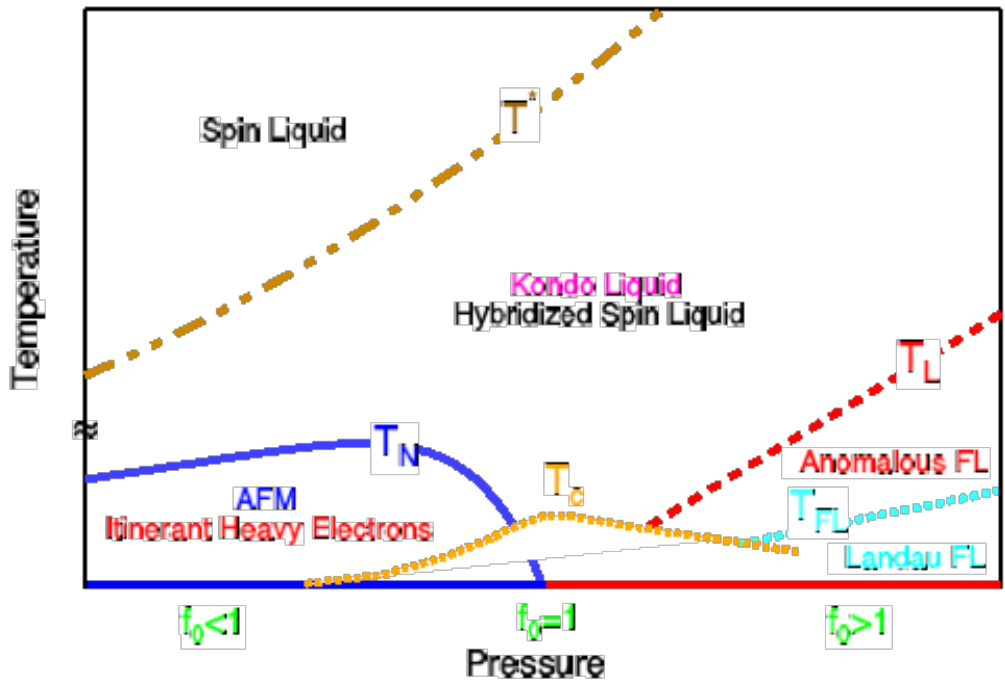

Fig. 5. A phase diagram for heavy electron materials in which the evolution with pressure of the hybridization effectiveness parameter, $f_0[P]$, determines the nature of the ordered ground state. Note that superconductivity emerges within the ordered antiferromagnetic state, that there is a quantum critical pressure and a QCP at $f_0=1$, and that for higher pressures there is a line, $T_L$, defined by the temperature at which $f_0=1$, below which no local moments exist and one has only a heavy fermion liquid whose behavior can be anomalous or reduce to that of a Landau Fermi liquid. In accord with the experimental results for both $CeRhIn_5$ and $CeCoIn_5$ under pressure, $T_c$ is shown as reaching its maximum value in the vicinity of this QCP. Figure from Yang and Pines (28).

Just as for the cuprates, spin forensics enables one to identify one component of the coexisting fluids below $T^*$; in this case it is the Kondo liquid that provides a key clue to magnetic behavior because its emergence can be seen as an anomaly when one plots the Knight shift against the uniform spin susceptibility, $\chi$. Above $T^*$, one has only one component, the unhybridized spin liquid, that reflects the mean field behavior of the local moment lattice; the Knight shift must then be proportional to its static spin susceptibility, $\chi_{SL}$. Below $T^*$, because a second component, the Kondo liquid, emerges, that proportionality no longer exists. However, one can use the measured Knight shift anomaly to track the emergent Kondo liquid [29] and show that its static susceptibility, $\chi_{KL}$, in the normal state follows the scaling law,

$$\chi_{KL} \sim f[T/T^*] [Rln2/2T^*] [2+ \ln (T^*/T)]; \qquad [7]$$

while Curro and his collaborators have shown that for all the 115 superconductors, the temperature dependence of $\chi_{KL}$ in the superconducting state is that expected for an unconventional d-wave superconductor [30].

The interplay between localization and itinerancy in the 115 materials and its impact on superconductivity is strikingly similar to that seen in the cuprates. As may be seen in the pressure-temperature phase diagram for $CeRhIn_5$ shown in Figure 6, the maximum superconducting transition temperature is found near the quantum critical pressure at which local moments emerge, while

Yang and the writer have shown that the emergence of superconductivity from 1 GPa to 1.75 GPa mirrors the increase in the fraction of heavy electrons able to participate in superconductivity that is measured by $f_{KL}(T_c)$ [28].

Further similarities become evident from a two-fluid analysis [31] of the pressure dependence of the spin-lattice relaxation rate in $CeCoIn_5$;[32]; in $CeCoIn_5$ the Kondo liquid begins to emerge at T*~48K, and becomes superconducting at 2.3K, close to the temperature at which the local moment contribution to system

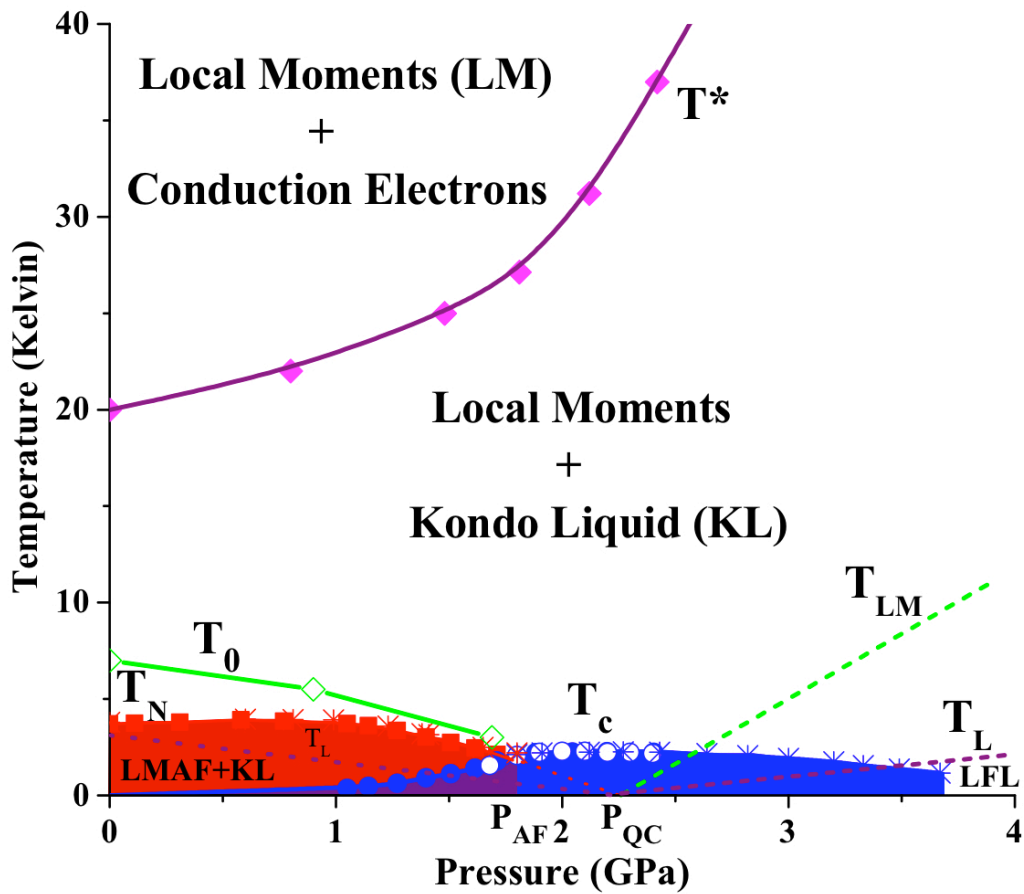

Fig. 6 The phase diagram for CeRhIn5 under pressure. Figure from Yang and Pines [28].

properties becomes vanishingly small. On writing

$$1/T_1 = [1-f(T/T^*)]/[T_1]_{SL} + f[T/T^*]/[T_1]_{KL}, \qquad [8]$$

assuming the hybridized spin liquid contribution below T* can be extrapolated smoothly from its behavior above T*, and making use of the two-fluid expression, Eq [5], for $f_{KL}$, Yang et al [31] obtain the results shown in Fig. 7. These demonstrate that the Kondo liquid spin-lattice relaxation rate in CeCoIn$_5$ exhibits a linear temperature dependence with a pressure dependent offset that is similar to the linear temperature dependence with a doping-dependent offset that we encountered in the local moment spin liquid in the underdoped cuprates, namely

$$[T_1]_{KL}T \sim [T+T_0(P)].$$

with $T_0$ increasing as the pressure increases. On extrapolating to negative pressure, one finds a QCP at ~ -0.5GPa. It is this QCP that is responsible for the spin fluctuations of the Kondo liquid that are identical in their impact on its spin-lattice relaxation rate to the impact of those produced by the low doping QCP in the cuprates on the $^{63}$Cu spin-lattice relaxation rate.

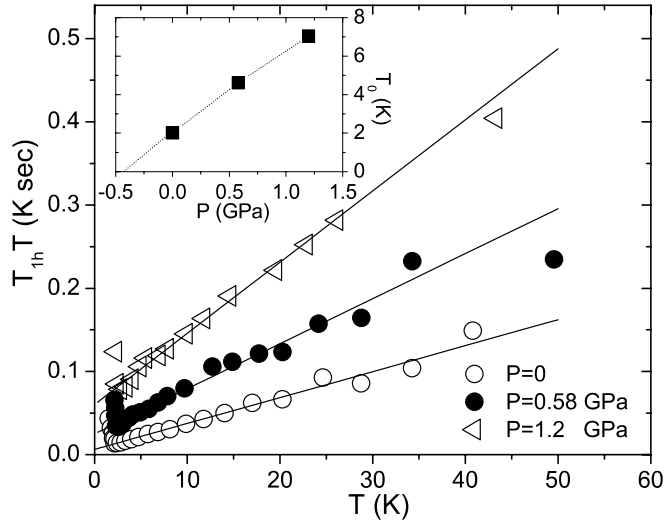

Figure 7. A two-fluid analysis of the NQR relaxation rate of CeCoIn$_5$ at different pressures [32] that reveals the linear in T behavior of the spin fluctuation spectrum of the Kondo liquid, with $T_1T \sim [T_0(P)+T]$ [31].

These striking similarities arise despite differences between the 115 heavy electron materials and the cuprates that include the physical origin of their local moments, the quite different temperature scales on which these hybridize, and the ordered state of their spin liquids. In the heavy electron materials, the local moments are an intrinsic property of the ions making up the lattice, while the low temperature state of their hybridized spin liquid is an antiferromagnet whose Neel temperature is reduced by the hybridization that occurs on an energy scale that is typically < 200K. In the cuprates, localization emerges from the strong Coulomb interaction between the planar orbital d electron states associated with the Cu ions, whose d-p hybridization with the orbital p states associated with the planar oxygen ions occurs on the ev scale, while the low temperature ordered state of the spin liquid that describes these local moments is the pseudogap state of matter. In the cuprates, since the planar d-p hybridization occurs on the ev scale, the order parameter describing the Fermi liquid that emerges is temperature independent by the time one reaches temperatures of order J, the nearest neighbor local moment coupling; in Kondo lattice materials the temperature scale, $T^*$, of the f-electron hybridization is $\sim J$.

The synthesis of a Pu-based "cousin" of the Ce-based 115 heavy electron family, $PuCoGa_5$, by Sarrao et al [33] and the subsequent measurements of its superconductivity by Curro et al [34] provide a dramatic example of the way in which the superconducting transition temperature can be enhanced within a given family. As we shall now see the enhancement can be traced to a six-to-seven-fold increase in the coupling J between their nearest neighbor local moments over that found in $CeCoIn_5$ .

The key step in establishing this connection was taken by Curro who found a quite remarkable correlation between the spin-lattice relaxation rate and superconducting transition temperature of $CeCoIn_5$ and $PuCoGa_5$ [34]. As may be seen in Figure 8, when scaled to their superconducting transition temperatures, he finds that the spin-lattice relaxation rate for the $PuCoGa_5$ material falls on that found for $CeCoIn_5$. Curro then went on to explore whether optimally doped YBa$_2$Cu$_3$O$_7$ displayed similar behavior and found that it does, thereby strengthening the argument that the spin fluctuations seen in the cuprate NMR experiments are the spin fluctuations responsible for their superconductivity.

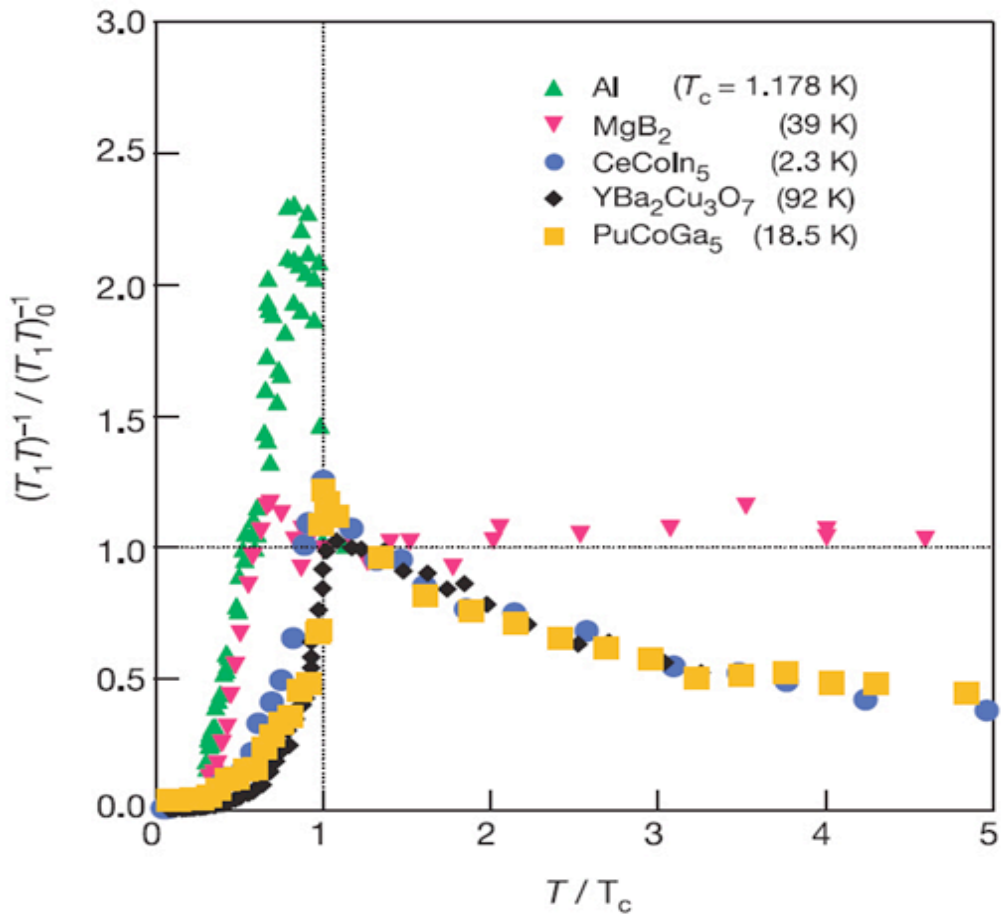

Figure 8. The spin lattice relaxation rate divided by temperature versus temperature for several key superconductors. The relaxation rate scale is normalized to account for different hyperfine couplings, and the temperature scale is normalized by the superconducting transition temperature. Al and MgB$_2$ are conventional s-wave superconductors well-described by the BCS theory, and exhibit a clear Hebel-Slichter peak below T$_c$ due to the coherence factors. On the other hand, CeCoIn$_5$, YBa$_2$Cu$_3$O$_7$, and PuCoGa$_5$ are unconventional superconductors. The striking feature is that the relaxation rate in the normal state of these compounds, which is dominated by antiferromagnetic spin fluctuations, scales with $T_c$. Figure reproduced from Curro et al.[34].

Curro's result is surprising for a number of reasons. First, in the case of $CeCoIn_5$ and $PuCoGa_5$ he is comparing their "unrenormalized" $T_1$ values; he has not subtracted the local moment contribution to focus on the "hidden" Kondo liquid spin lattice relaxation rate produced by the quantum critical spin fluctuations that bring about their superconductivity. Second, in comparing their behavior to that of $YBa_2Cu_3O_7$, he would seem to be comparing apples and oranges, in that the $YBa_2Cu_3O_7$ spin-lattice relaxation rate is produced by the low frequency fluctuations of the still-present spin liquid, while for heavy electron materials one gets contributions to $T_1$ from both the spin and Kondo liquids.

Although the explanation for the first surprise is straightforward, it has not been noted previously. First, since $T^*$ is the only characteristic temperature that enters into the spin-lattice relaxation rate, we should first ask whether it exhibits $T^*$ scaling behavior; Curro [35] has now found that it does. Second, from Eq. 8 we see that the local moment and Kondo liquid contributions must then separately scale with $T^*$, and that for the spin liquid such scaling can occur only at the quantum critical pressure where $f_0=1$. Third, for the Kondo liquid contribution to scale with $T^*$, we see from Eq [7] that the offset, $T_0(P_c)$ must either be proportional to $T^*$ or that the scaling results are not sensitive to its value.

But, as we have seen in Fig.8, the total spin lattice relaxation rate also scales with $T_c$. This is only possible if $T_c$ scales with $T^*$, so what the Curro results are telling us is that at the quantum critical pressure at which $f_0=1$, $T_c$ will be maximum and will scale with the nearest neighbor local moment interaction, J. Moreover, since the nearly optimally doped member of the 123 family displays similar scaling behavior down to $T^*$, something comparable might be expected for the optimally doped members of the cuprate family.

However, for the cuprates, the argument is not straightforward, because what is being measured are the spin fluctuations of the spin liquid, which does not directly participate in their superconductivity, so for the spin liquid itself, one is initially interested in their connection to the onset of its pseudogap behavior.

So let us ask whether there is a doping concentration for the cuprates at which one gets comparable $T^*$ scaling of their spin-lattice relaxation rate? The answer is yes, and somewhat surprisingly, it turns out to be at a doping level quite close to that at which $T_c$ is maximum. The algebra is straightforward:

$$[T_1T/T_1T^*] = [T_0(x)/T^*(x) + T/T^*(x)] / [T_0(x)/T^*(x) + 1] \quad [9]$$

which, upon making use of $T^* = J_{eff}/3$, Eqs [3] and [4], and requiring that one get the same scaling result as found by Curro for the heavy electron materials (35),

$$[T_1T/T_1T^*] = [1 + T/T^*]/2$$

leads to the following results for the 123 and 214 cuprates,

123 :   $x_c = 1.0145$ and $T_0 = T^* = 94K$
214:   $p_c = 0.165$ and   $T_0 = T^* = 92K$

The quantum critical doping levels, $x_c$ and $p_c$, that yield cuprate scaling similar to that found in the heavy electron materials turn out to be remarkably close to those at which $T_c$ is optimal. This demonstrates a striking similarity for optimally doped materials between the build up of the spin fluctuations that bring about pseudogap behavior in the spin liquid and those responsible for superconductivity in the quasiparticle liquid, and further implicates the spin liquid in bringing about superconductivity in optimally doped cuprates.

Our conclusions from the scaling results are summarized in Table 1. We see that in assuming $T_c/J$ scaling for the cuprates we come reasonably close: the measured $T_c$ for $La_{1.835}Sr_{0.165}CuO_4$ is some 8K lower than simple scaling with heavy electron materials would predict, while that for $YBa_2Cu_3O_7$ is some 35K larger.

Table 1.  A comparison of optimal superconducting transition temperatures for cuprate and 115 materials with the strength of their nearest neighbor local moment interaction. For the cuprates J is taken from fits of the 2d Heisenberg model to the antiferromagnetic state at zero doping; for $CeCoIn_5$ it is taken from estimates of $T^*$ at 1.3GPa, while for $PuCoGa_5$, since no direct measurements of $T^*$ exist, J is an estimate based on fits to lower temperature data. Should it turn out to be ~490K, $T_c/J$ would be the same as that found for optimally pressurized $CeCoIn_5$.

| Material | J[K] | $T_c$[K] | $T_c/J$ |
| --- | --- | --- | --- |
| $CeCoIn_5$ at 1.3GPa | ~70 | 2.6 | ~0.037 |
| $PuCoGa_5$ at svp | ~400 | 18.4 | ~0.046 |
| $La_{1.84}Sr_{.16}CuO_4$ | ~1290 | 40 | ~0.031 |
| $YBa_2Cu_3O_7$ | ~1500 | 90 | ~0.060 |

Conclusion and Acknowledgments

　　We have seen how experiment combined with phenomenology and theoretical models tells us how to use the spin-fluctuation gateway to finding higher temperature superconductors: find a material with a Neel temperature considerably greater than 400K in which nearly two-dimensional superconductivity emerges when its properties are changed by doping and/or pressure. Since for such materials it appears reasonable to assume that $T_c \sim J/25$, and since materials with Neel temperatures in excess of 1000K have already been discovered [$SrTcO_3$,MnNi,…], on further assuming that for some of these J could be as large as 4000K to 5000K, there would not seem to be a firm magnetic ceiling to stop one from coming close to room temperature superconductivity, provided one finds materials with such large J's that contain planes that can conveniently be doped to produce superconductivity.

　　It gives me great pleasure to dedicate this paper to Peter Wolynes, a long-time friend, and a past and, I hope, future collaborator on designing emergent matter, on the occasion of a 60[th] birthday that marks only the start of the many further decades of significant scientific accomplishment we hope and expect for him. I should like to thank my present collaborators in research on unconventional superconductors, Nick Curro, Zachary Fisk, Gil Lonzarich, and Yi-feng Yang, for many helpful discussions on these and related topics, and to acknowledge how much I have learned during the past three decades from the members of my Urbana research group--my then graduate students, Victor Barzykin and Philippe Monthoux, and my then postdocs, Sasha Balatsky, Andrey Chubukov, Hartmut Monien, Dirk Morr, Joerg Schmalian and Branko Stojkovic-- and my collaborators there and elsewhere, Andy Millis, Chris Pethick, Charlie Slichter, and Joe Thompson. In addition I thank Nick Curro and Yi-feng Yang for their help in preparing some of the figures that appear here, Elihu Abrahams, Zachary Fisk and Yi-feng Yang for a critical reading of early versions of this manuscript, and the Santa Fe Institute for its hospitality during its preparation.

Footnotes